\documentclass[]{emulateapj}
\usepackage[usenames]{color}

\newlength{\figwidth}
\def\lsim{\mathrel{\rlap{\lower4pt\hbox{\hskip1pt$\sim$}}
    \raise1pt\hbox{$<$}}}                

\begin{document}

\title{The Scalar Perturbation Spectral index \lowercase{$n_s$}: \emph{WMAP} sensitivity to unresolved point sources}


\author{
K.~M.~Huffenberger,\altaffilmark{1,2,3}
H.~K.~Eriksen,\altaffilmark{4,5} 
F.~K.~Hansen,\altaffilmark{4,5}
A.~J.~Banday,\altaffilmark{6}
K.~M.~G\'orski\altaffilmark{2,3,7}
}

\altaffiltext{1}{huffenbe@jpl.nasa.gov}
\altaffiltext{2}{Jet Propulsion Laboratory, 4800 Oak  Grove Drive, Pasadena CA 91109} 
\altaffiltext{3}{California Institute of Technology, Pasadena, CA  91125} 
\altaffiltext{4}{Institute of Theoretical Astrophysics, University of Oslo, P.O.\ Box 1029 Blindern, N-0315 Oslo, Norway}
\altaffiltext{5}{Centre of Mathematics for Applications, University of Oslo, P.O.\ Box 1053 Blindern, N-0316 Oslo}
\altaffiltext{6}{Max-Planck-Institut f\"ur Astrophysik,
  Karl-Schwarzschild-Str.~1, Postfach 1317, D-85741 Garching bei M\"unchen, Germany}
\altaffiltext{7}{Warsaw University Observatory, Aleje Ujazdowskie 4, 00-478 Warszawa, Poland}

\date{Received - / Accepted -}


\begin{abstract}
  Precision measurement of the scalar perturbation spectral index,
  $n_s$, from the \emph{Wilkinson Microwave Anisotropy Probe}
  temperature angular power spectrum requires the subtraction of
  unresolved point source power. 
  Here we reconsider this
  issue, attempting to resolve inconsistencies found in
  the literature. 
  First, we note a peculiarity in the \emph{WMAP} temperature likelihood's
  response to the source correction: 
  Cosmological parameters do not respond to increased source errors.  
  An alternative and more direct
  method for treating this error term acts more sensibly,
  and also shifts $n_s$ by $\sim0.3\sigma$ closer to unity.
  Second, we re-examine the source fit used to correct the power
  spectrum.  This fit depends strongly on the galactic
  cut and the weighting of the map, indicating that either the source
  population or masking procedure is not isotropic.  
  Jackknife
  tests appear inconsistent, causing us to
  assign large uncertainties to account for
  possible systematics. 
  Third, we note that the \emph{WMAP} team's spectrum
  was computed with two different weighting
  schemes: uniform weights transition to inverse noise
  variance weights at $l=500$.  The fit depends on such weighting schemes, so
  different corrections apply to each multipole range.
  For the
  Kp2 mask used in cosmological analysis, we prefer source corrections 
  {$A=0.012\pm0.005$ $\mu$K$^2$} for uniform weighting and 
  {$A=0.015\pm0.005$ $\mu$K$^2$} for $N_{\rm obs}$ weighting. Correcting \emph{WMAP}'s spectrum
  correspondingly, we compute cosmological parameters with our
  alternative likelihood, finding 
  $n_s=0.970\pm0.017$ 
  and
  $\sigma_8=0.778\pm0.045$.  
  This $n_s$ is only $1.8\sigma$ from unity, compared to  the  $\sim 2.6\sigma$
  \emph{WMAP} 3-year result.
  Finally, an anomalous feature in the
  source spectrum at $l<200$ remains, most strongly associated with W-band. 
\end{abstract}
\keywords{cosmology: observations, cosmic microwave background,
  cosmological parameters, methods: data analysis}

\section{Introduction}

Measuring $n_s$, the spectral index of initial scalar fluctuations,
which is scale invariant ($n_s = 1$) in the Harrison-Zeldovich model
and slightly shallower in inflation models, is difficult, primarily
because experimental systematics require control over a broad range of
spatial scales.  In inflation, the deviation from unity closely
relates to the inflationary potential and the number of $e$-folds of
expansion, so a statistically robust measurement of $n_s \neq 1$
places compelling constraints on the physics of the inflationary
epoch.

Because all-sky measurements of the cosmic microwave background (CMB)
access the largest observable scales in the universe, the angular
power spectrum of the CMB, with a long lever arm, is crucial to such
studies.  Indeed, the latest data release from the \textit{Wilkinson
  Microwave Anisotropy Probe} (\emph{WMAP}) claims $\sim 2.6\sigma$
deviation from the Harrison-Zeldovich spectrum \citep{Spergel2007}.
Unfortunately, the CMB is not a totally clean measurement.  For
example, the well-known degeneracy with the optical depth since
recombination ($\tau$) makes precision measurement of $n_s$ impossible
using CMB temperature anisotropies alone, and polarization is required
to break it. Complicated noise properties and hints of unknown
systematics in the \emph{WMAP} measurement of large-scale polarization
indicate that the systematic
uncertainty in both $\tau$ and $n_s$ should still be considered
significant \citep{Eriksen2007b}.

Another important, but under-appreciated, complication for the
measurement of $n_s$ is additional power in the angular spectrum from
unresolved, and unmasked, point sources.  At high $l$, this shot noise
can significantly bias the power spectrum, and consequently $n_s$.
The \emph{WMAP} team devised a sensible prescription for dealing with
this contaminant: 1) Use the spectral energy distribution measured
from detected sources (and distinct from the CMB) to infer it for
undetected ones; 2) measure the contamination using multifrequency
data; 3) correct the spectrum; and 4) marginalize over the measurement
error when computing the likelihood \citep{Hinshaw2003,Hinshaw2007}.

\citet{Huffenberger2004} found a level of source contamination
consistent with the level in the first \emph{WMAP} data release
\citep{Hinshaw2003}.  However, based on the three year temperature
data \citep{Hinshaw2007}, \citet{Huffenberger2006} measured a point
source spectrum with two irregularities.  First, at $l>200$ the
spectrum is white, but with an amplitude below the value in the
original preprint of \citet{Hinshaw2007}. In the present work, we
discovered a small error in the power spectra used for the
\citet{Huffenberger2006} estimate, which should have reported $A =
0.013 \pm 0.001$ $\mu$K$^2$ instead of $A = 0.011 \pm 0.001$
$\mu$K$^2$, still below the original WMAP value of $A = 0.017 \pm
0.002$ $\mu$K$^2$.  Prompted by our result, \citet{Hinshaw2007}
re-examined the issue, revising their value down somewhat and
increasing the error bars, to $A = 0.014 \pm 0.003$ $\mu$K$^2$.  The
\citet{Spergel2007} bispectrum analysis indicates a non-Gaussianity
consistent with these values, but lacks the statistical power of the
multifrequency power spectrum comparison.  The second peculiarity is
that the power at $100<l<200$ in \citet{Huffenberger2006} was
inconsistent, at strong statistical significance, with the rest of the
white spectrum.

This paper again considers the power spectrum source correction
procedure in detail.  We begin in Section \ref{sec:ns_impact} with a
study of the impact of the source correction on the scalar spectral
index through the likelihood.  Following this, we probe the source
amplitude in Section \ref{sec:source_spectra}, examining the
dependence of the fit on the sky weighting, mask, year of observation,
and frequency dependence, and present our best estimates of the
cosmological parameters.  These same tests probe the robustness of the
$l<200$ feature.  Finally, we conclude in Section
\ref{sec:conclusions}.

\section{Source correction impact on spectral index}
\label{sec:ns_impact}

The final \emph{WMAP} temperature power spectrum is a noise-weighted
combination of cross spectra computed from V- (61 GHz) and W-band (94
GHz) maps.  Prior to the computation of the angular spectrum, a
foreground model is removed from the maps.  The individual
cross-spectra are corrected for the sky mask, instrument beams, and
point source contamination before combination
\citep[see][]{Hinshaw2007}.  The combined spectrum is folded into the
likelihood calculation,\footnote{We are using version 2.2.2 of the
  \emph{WMAP} likelihood, available at
  \texttt{http://lambda.gsfc.nasa.gov/}.} which interfaces to a Markov Chain Monte Carlo code such
as \emph{CosmoMC} \citep{LewisBridle2002}, yielding parameter
estimates.

\begin{figure}
  \includegraphics[width=\figwidth]{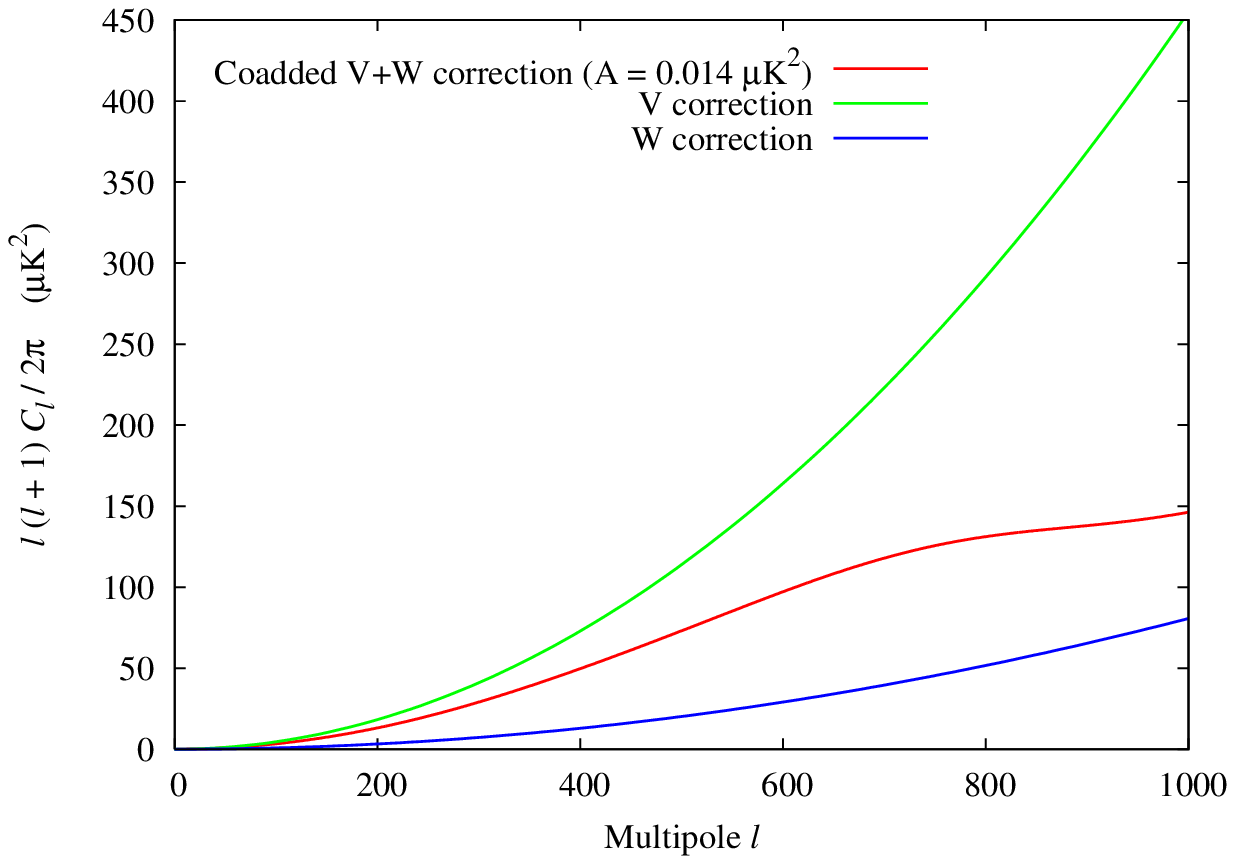}
\caption{The size of the source correction in \citet{Hinshaw2007}.} \label{fig:src_corr}
\end{figure}

Assuming the angular spectrum for sources is white, we wish to explore
the dependence of $n_s$ on the size of the source correction.  In
Figure \ref{fig:src_corr}, we show the correction made by
\citet{Hinshaw2007}.  The correction is not white, because the
combined spectrum gets more contribution from W-band at higher $l$, due to
the noise weighting.  To change the amplitude of the correction, we
simply scale.  Later, we discuss the estimation of this amplitude.

We explore two methods for incorporating the source errors into the
likelihood.  In the first, we use the standard \emph{WMAP} likelihood
method.  The total log-likelihood is split in two, $-2 \log L = {\cal
  L} = {\cal L}_0 + {\cal L}_1$, where ${\cal L}_0$ is due to cosmic
variance, noise, and mask-induced mode coupling, and ${\cal L}_1$
includes the uncertainty due to the point source and beam errors
\citep{Hinshaw2003,Hinshaw2007}.  The computation of ${\cal L}_1$
assumes that total likelihood is Gaussian, and uses the Woodbury
formula to compute an update to the likelihood.  Under the Gaussian
assumption this split is exact.  However, the ${\cal L}_0$ term is not
treated as Gaussian, but computed using the Gaussian plus log-normal
approximation \citep{Verde2003}.

Treating the source and beam uncertainty separately is very fast
(since the uncertainty can be well approximated by a small number of
modes), but not really necessary.  It avoids the inversion of a $\sim
1000 \times 1000$ matrix per likelihood evaluation, but the low-$l$
part of the code already inverts a $\sim 2000 \times 2000$ matrix.
Performing this inversion and including the beam and source term with
the other sources of error in ${\cal L}_0$ is therefore little
additional burden.

We take this step in our second method for incorporating the source
errors.  We have modified the \emph{WMAP} likelihood code to integrate
the beam/point source covariance matrix into the cosmic
variance/noise/mask covariance matrix.  (This requires inverting the
original Fisher matrix, adding the beam/point source term, and
inverting back.)  Then we feed the new covariance matrix to the
Gaussian plus log-normal approximation.  Under this procedure, the
change in the likelihood due to the inclusion of the beam/point
sources term is -2.64, compared to ${\cal L}_1 = -1.22$ computed with
the Woodbury expansion, for the test theory spectrum included with the
\emph{WMAP} likelihood code (where ${\cal L} \sim 3541$).

\begin{figure}
  \begin{center}
    \includegraphics[width=\figwidth]{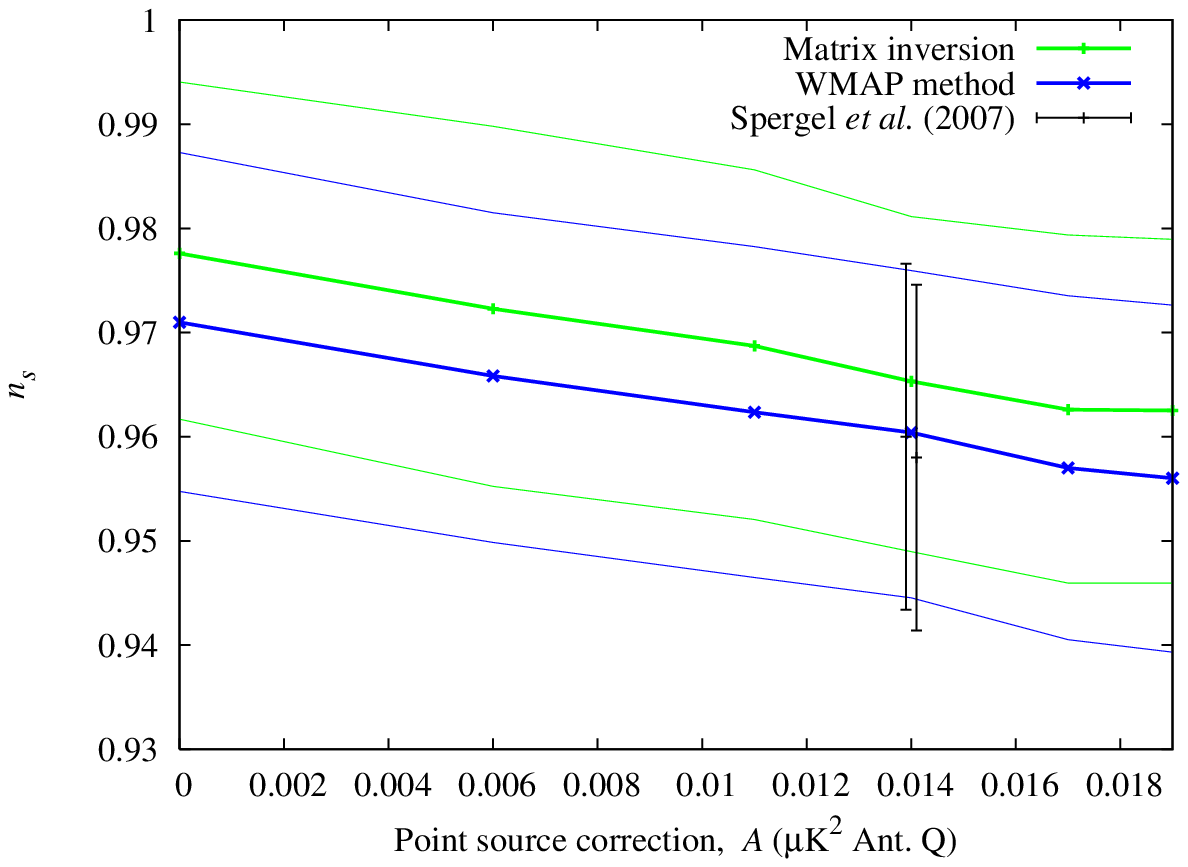}
    \includegraphics[width=\figwidth]{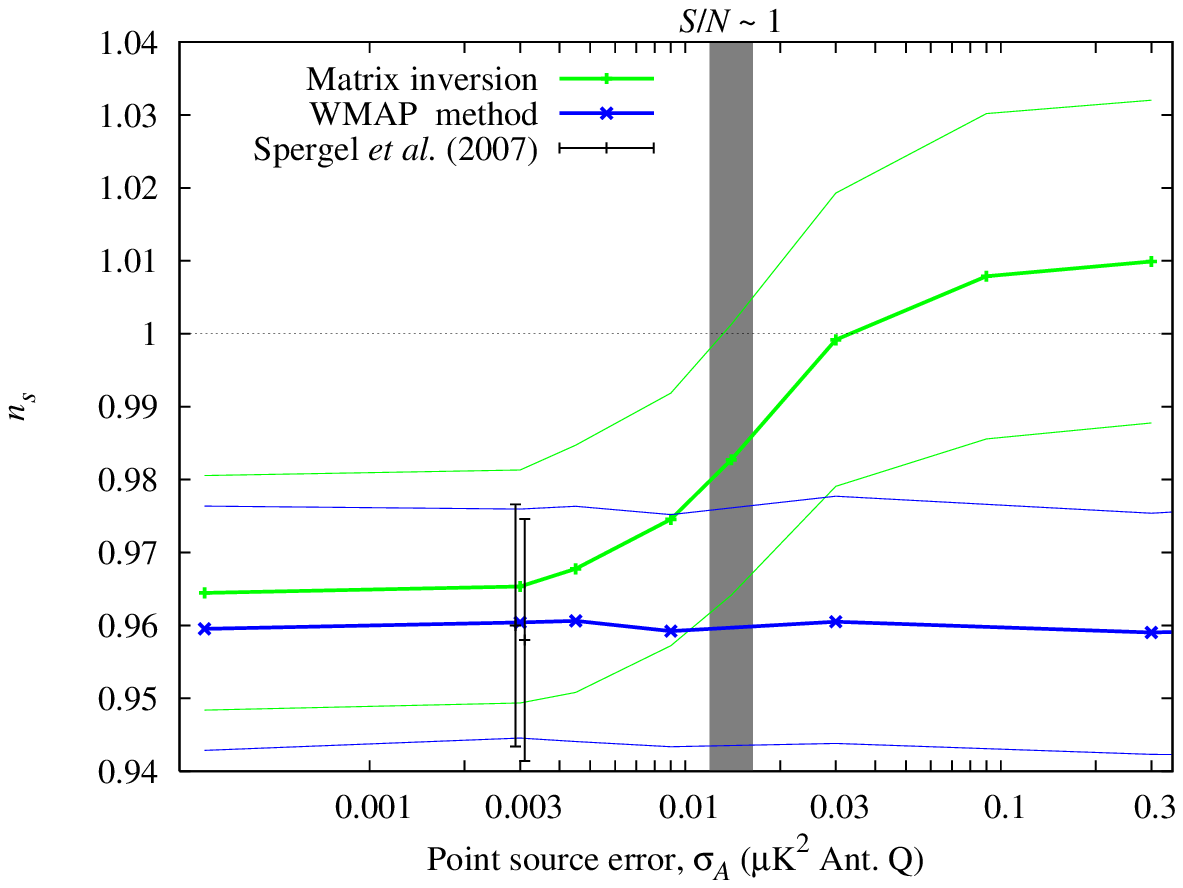}
  \end{center}
  \caption{(Top) The change in $n_s$ for given source corrections,
    $A$.  We compare the \emph{WMAP} team's likelihood code to our
    modified version, which computes the high-$l$ likelihood in a
    slightly different way.  The error in the source correction is
    fixed at the \citet{Hinshaw2007} value used in
    \citet{Spergel2007}.  The thin lines bound the 68\% probability
    interval.  The \citet{Spergel2007} values are slightly offset
    horizontally for visibility, and either ignore (higher $n_s$) or
    include (lower) a correction for Sunyaev-Zeldovich effect
    contamination.  The other $n_s$ values ignore SZ.  (Bottom) Using
    the \emph{WMAP} team's source correction, this plot shows the
    dependence of $n_s$ on the point source error, $\sigma_A$.  }
  \label{fig:ns_dep}
\end{figure}
In the top panel of Figure \ref{fig:ns_dep}, we show the dependence of
$n_s$ on the source correction.  We hold the errors fixed at the
\citet{Hinshaw2007} value of $\sigma_A = 0.003$ $\mu$K$^2$, and
marginalize over all the other parameters.  As the source correction
increases, the power spectrum at high $l$ is lowered, and $n_s$
decreases.  Completely ignoring the source correction shifts $n_s$ by
$\sim 0.01$ higher, or about $0.6\sigma$.  Our alternative likelihood
procedure shifts $n_s$ higher by $0.005$, or $0.3\sigma$.

Next we study the influence of error in the source correction.  At
very small and at very large error, we expect the parameter
measurement to be independent of the error.  For very small source
error, the errors in other quantities dominate.  For large errors, all
modes which could be contaminated by point sources are effectively
projected out, and the parameter measurement is again independent.
Near $S/N \sim 1$, the measurement should undergo a transition, where
the measurement error increases and (possibly) the mean value changes.
In the bottom panel of Figure \ref{fig:ns_dep}, we show the dependence
of $n_s$ on the error on the source correction amplitude.  The
\emph{WMAP} split likelihood method shows an unexpected result.  The
size of the point source error apparently has no effect on the
measurement of $n_s$.  Even with the point source error rivaling the
size of the acoustic peaks, there is no effect.  The values of the
likelihood change, but not the distribution of points in the Markov
Chain.  This seems to be a clear indication that there is something
wrong.  The dependence of $n_s$ on $A$ in the top panel implies that
at least the error bars should increase as $\sigma_A$ increases.

On the other hand, our modified likelihood shows the expected
behavior.  As the source correction error is made very large, the
errors increase by about 38 percent, and the mean value moves above 1.
For this likelihood, the modes not subject to contamination by sources
actually prefer $n_s > 1$, and a solid measurement of the source
contamination is vital to the measurement of $n_s < 1$.  We also note
that the errors on the source measurement do not make much difference,
as long as $\sigma_A < 0.003$.  In the next Section, we examine this
measurement in more detail.

\section{Unresolved point source spectra} \label{sec:source_spectra}

\subsection{Method}
\label{sec:method}
The point source spectrum can be estimated via a linear combination of
the individual cross-spectra at several frequencies, a combination
which projects out the CMB component.
\citet{Huffenberger2004,Huffenberger2006} examined the unresolved
source component in \emph{WMAP} data, using a generalized version of
the method for the same task from \citet{Hinshaw2003}.

In addition to V- and W-band used for cosmological measurement, the
source analysis uses Q-band (41 GHz) because the contaminating sources
are much brighter at lower frequencies.  There are 276 cross-spectra
in these three bands, accounting for all combinations of differencing
assemblies per channel, and treating the three years of observation
separately.  These are combined as
\begin{equation}
 A_L = \sum_{L'} \sum_{i \neq j} W_{LL'}^{ij} D_{L'}^{ij}
\label{eq:amplitude}
\end{equation}
to give the point source amplitude $A_L$ in a band denoted by $L$,
where the weight is $W_L^{ij}$ for the binned cross spectrum estimate
$ D_L^{ij}$, made from maps $i$ and $j$.  The map cross spectra
$D_L^{ij}$ include the CMB power spectrum and the contribution from
sources.  By virtue of being cross-spectra ($i \neq j$), they are
noisy but do not have a noise bias.  The weights are based upon the
frequency independence of the CMB signal (in thermodynamic temperature
units), the spectral energy distribution of the sources (measured for
bright sources as $\beta \sim -2.0$, $S \propto \nu^{\beta+2}$
\citep{Hinshaw2003,Trushkin2003,Hinshaw2007}), and the estimated noise
covariance in the cross spectrum measurements.  (See
\citet{Huffenberger2004,Huffenberger2006} for details.)  

The weights obey the constraint
\begin{equation}
  \sum_{L'} \sum_{ij} W_{LL'}^{ij} = 0,
\end{equation}
which means that no CMB will leak into the point source estimate if
the maps are properly calibrated and the instrumental beams are
perfectly deconvolved from the spectra.  If the source spectral energy
distribution is correct, the weights also provide an unbiased estimate
of the point source power spectrum even if the noise covariance is
wrong (although an incorrect noise covariance leads to sub-optimal
estimates and incorrect error bars).  

\begin{figure}
\begin{center}
\includegraphics[width=\figwidth]{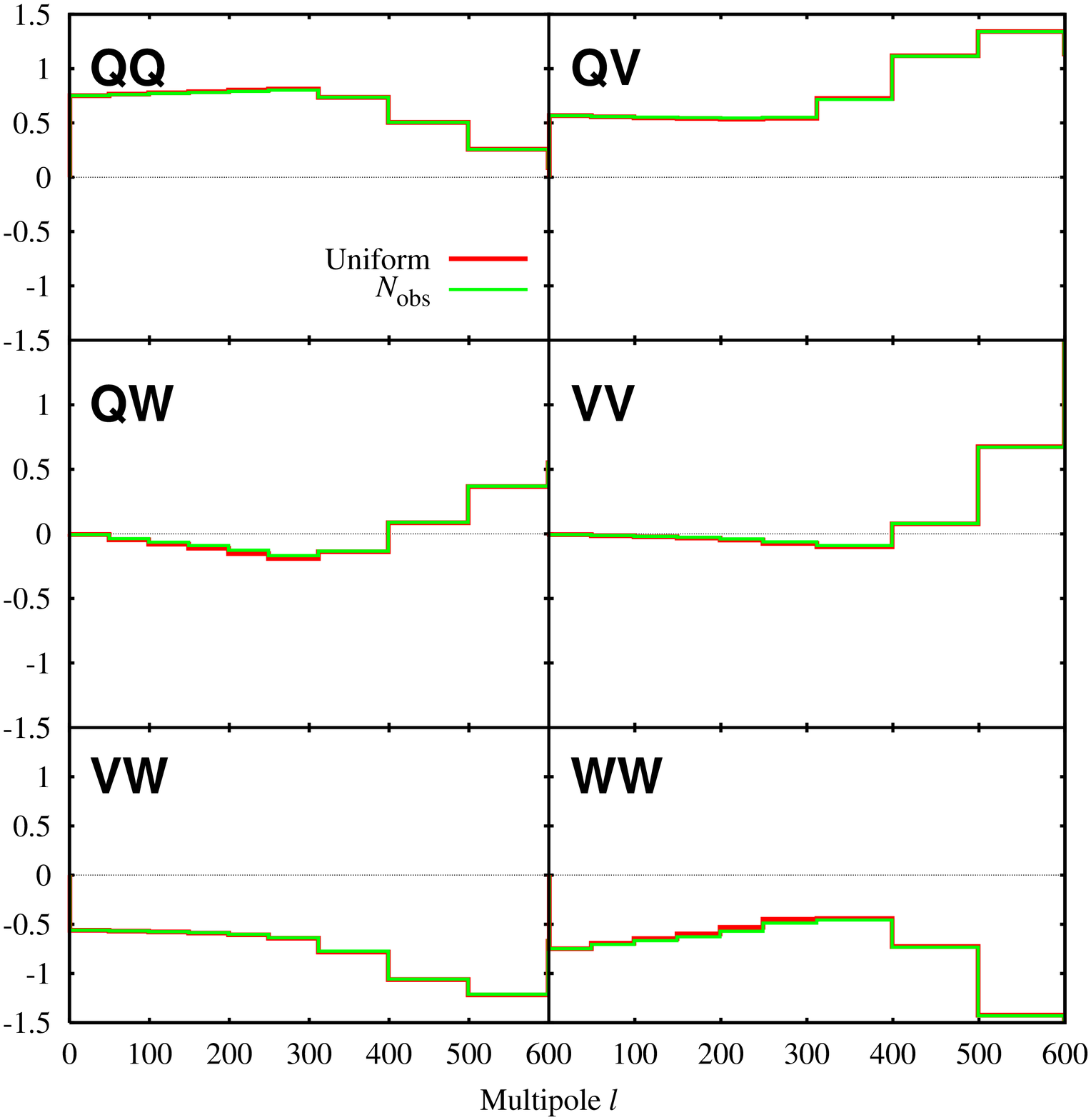}
\end{center}
\caption{The weights for various cross-spectra in the source measurement, using the Kp2 mask, and two different map weighting.  The individual weights for the 276 cross-spectra were summed together based on frequency coverage, yielding the 6 combinations shown.}
\label{fig:weights}
\end{figure}

The covariance of the source power spectrum estimate is
\begin{equation}
  \langle  A_L A_{L'}   \rangle = ({\mathbf {W \Sigma_D W }})_{LL'},
\end{equation}
where $\mathbf {\Sigma_D}$ is the estimate of the cross spectrum
covariance matrix.  If the covariance matrix is diagonal in
cross-spectra and multipole bin, then the weights are diagonal in
multipole.  Under this assumption, we plot some example weights in
Figure \ref{fig:weights}.

The expected shot noise angular spectrum of sources in the \emph{WMAP}
data is flat in $C_l$, so we plot our measured spectra as $l$ versus
$C_l$.  Throughout we normalize the spectrum at Q band in antenna
temperature, which gets a larger signal than V or W. 

In the following subsections, we describe several tests of the point
source measurement to explore the robustness of these features and
their origin.  Tests included changing the weights on the map for
computing the spectrum, modifying the mask,
fitting for the galactic hemispheres separately, fitting year by year,
and changing the assumed source spectral energy distribution.


\subsection{Spectrum at $l > 200$}

Whatever the cause of the excess power at $l<200$ reported by
\citet{Huffenberger2006} may be, it is very likely not relevant for
correcting the spectrum at high $l$. In this Section, we therefore
first concentrate on the correction required for the cosmological
analysis, considering only $l \ge 200$, and then return to the
anomalous low-$l$ feature in Section \ref{sec:lowl_excess}.

In a maximum likelihood estimate of the power spectrum, the map is
weighted by the pixel-pixel inverse signal plus noise covariance.  At
small scales this estimate is computationally impractical, and
\citet{Hinshaw2007} instead approximate it using two weighting schemes
in the computation of the power spectra.  In the signal dominated
regime at $l<500$, they use spectra where every pixel is weighted
evenly; in the noise dominated regime at $l>500$, the maps are inverse
noise weighted (i.e.  weighted by the number of observations, $N_{\rm
  obs}$).

\begin{figure}
\begin{center}
  \includegraphics[width=\figwidth]{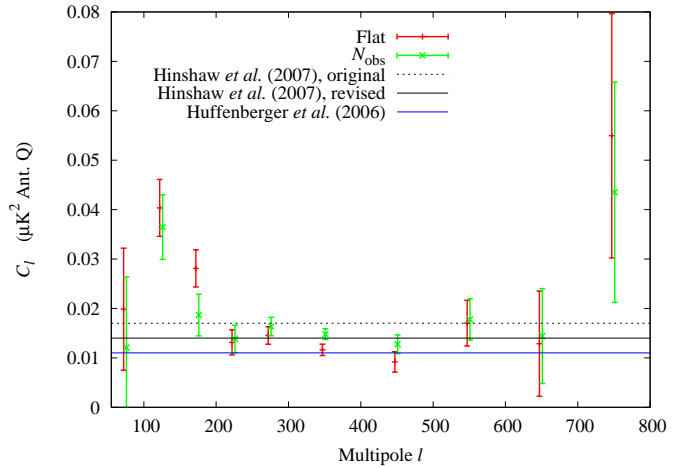}
\end{center}
\caption{Fits for the unmasked source component, comparing spectra computed from maps using uniform weight outside of the mask with using $N_{\rm obs}$ (inverse variance) weighting, plotted as the Q-band amplitude.  The lowest $l$-bin has no detection and is not shown.}
\label{fig:nobs}
\end{figure}

We compute the cross-spectrum using these two schemes and estimate the
source contribution, plotted in Figure \ref{fig:nobs}.
The flat portion of the spectrum is notably higher in the $N_{\rm
  obs}$ weighted case.  Specifically, for a white noise fit including
$l>200$, $A = 0.015 \pm 0.001$ $\mu$K$^2$ versus $A = 0.012 \pm 0.001$
$\mu$K$^2$ for the flat weighting.  (The single amplitude fits for all
combinations we tried are compiled in Table~\ref{tab:fits}.)  This
leads immediately to a key point: Since the combined spectrum is built
from two weighting schemes at different $l$'s, the source correction
must be different as well.  Thus we should have a smaller source
correction for $l<500$ and a larger correction for $l>500$.  The
weighting scheme used by \citet{Hinshaw2007} is not specified.

\begin{deluxetable}{lllccc}
\tabletypesize{\small} 
\tablecaption{Fits for source amplitude \label{tab:fits}} 
\tablecolumns{6}
\tablehead{Mask&Weight&Subset&$A$&$\sigma_A$ & $\chi^2$/dof\\ & & &\multicolumn{2}{c}{($\mu$K$^2$, ant. Q)} &}

\tablecomments{Fits for source power spectrum amplitude, including data at $l>200$.}

\startdata

\cutinhead{Diagonal Covariance}

30kp0 & flat & all & 0.006 & 0.001 & 0.25 \\
kp2 & flat & VW & 0.006 & 0.004 & 1.99 \\
kp0 & flat & all & 0.008 & 0.001 & 0.54 \\
kp2 & flat & noQ1 & 0.010 & 0.001 & 2.02 \\
kp2 & flat & noQ2 & 0.010 & 0.001 & 0.71 \\
kp2 & flat & yr12 & 0.011 & 0.002 & 1.02 \\
kp2 & flat & yr13 & 0.011 & 0.002 & 1.10 \\
kp2 & flat & all (N. hemi.) & 0.011 & 0.001 & 0.92 \\
kp2 & flat & all & 0.012 & 0.001 & 1.39 \\
kp2 & flat & yr22 & 0.012 & 0.003 & 1.16 \\
kp2 & flat & yr23 & 0.012 & 0.002 & 0.50 \\
kp2 & flat & QW & 0.013 & 0.001 & 0.70 \\
kp2 & flat & all (S. hemi.) & 0.013 & 0.001 & 0.68 \\
kp2 & flat & noW1W2 & 0.013 & 0.001 & 2.48 \\
kp2 & flat & noW3W4 & 0.013 & 0.001 & 1.23 \\
kp2 & flat & QV & 0.014 & 0.001 & 1.65 \\
kp2 & nobs & all & 0.015 & 0.001 & 0.69 \\
kp2 & flat & yr11 & 0.017 & 0.003 & 0.87 \\
kp2 & flat & yr33 & 0.017 & 0.003 & 0.20 \\

\cutinhead{Diagonal + beams errors}

kp2 & flat & noQ1 & 0.009 & 0.001 & 1.87 \\
kp2 & flat & VW & 0.010 & 0.004 & 2.07 \\
kp2 & flat & noQ2 & 0.010 & 0.001 & 0.73 \\
kp2 & flat & yr12 & 0.011 & 0.002 & 1.05 \\
kp2 & flat & yr13 & 0.011 & 0.002 & 1.10 \\
kp2 & flat & yr22 & 0.011 & 0.003 & 1.11 \\
kp2 & flat & yr23 & 0.011 & 0.002 & 0.34 \\
kp2 & flat & all & 0.012 & 0.001 & 1.53 \\
kp2 & flat & QW & 0.013 & 0.001 & 0.72 \\
kp2 & nobs & all & 0.015 & 0.001 & 0.79 \\
kp2 & flat & QV & 0.016 & 0.001 & 2.30 \\
kp2 & flat & yr33 & 0.017 & 0.003 & 0.25 \\
kp2 & flat & yr11 & 0.019 & 0.003 & 0.85 \\

\enddata

\end{deluxetable}

This source level difference is a strong indication that the source
population contaminating the \emph{WMAP} spectrum is not isotropic.
One possible scenario is the following: Because source positions are
stochastic, and source power strongly favors brighter sources ($C_l
\propto \int dS\, S^2 \, dN/dS$), brighter sources which, by chance,
fell in the best observed regions could boost the spectrum in the $N_{\rm
  obs}$ weighting over uniform weighting. However, the size of this
effect is much too small to be a viable explanation.   We computed power
spectra of Monte Carlo realizations of a noiseless map containing
isotropically distributed faint sources, based on differential source
counts $dN/dS \propto S^{-2.3}$ \citep{WhiteMaj2004,Cleary2005},
normalized to \emph{WMAP} source counts at 1 Jy, setting an upper flux
limit to reproduce a reasonable amount of power.  Over 1000
simulations, for a bin $100 < l < 150$, the rms fluctuation of the
difference in power between the uniform and $N_{\rm obs}$ spectra is
1.8 percent, and should be smaller at higher $l$.  Thus the odds are
very slight that the change in the power is due to chance alignments
of sources with the well observed part of the sky; sources at the
appropriate flux level are too numerous.  Alternatively, the observed
anisotropy in the source population could be either be representative
of the real sources, or indicate a problem with the masking procedure,
with bright sources slipping through.

The $N_{\rm obs}$ weighting emphasizes the ecliptic poles, so these
are the best places to look for suspect sources.
\begin{figure}
\begin{center}
\includegraphics[width=0.95\figwidth]{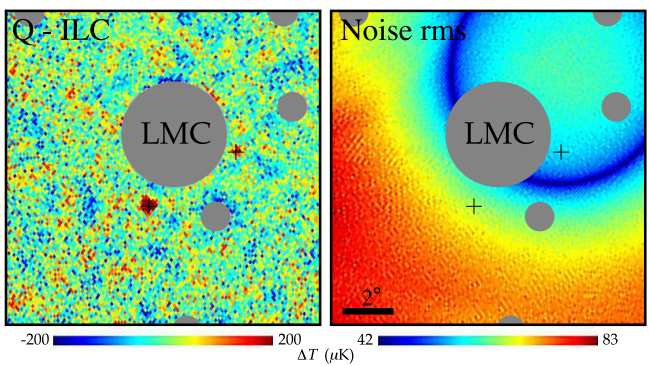}
\end{center}
\caption{The difference map, Q$-$ILC, shows two point-like objects near the LMC, left, compared to the Q-band noise rms for the same region, right, based on the number of observations reported by the \emph{WMAP} team.}
\label{fig:lmc}
\end{figure}
In particular, visual inspection of a Q$-$ILC difference map
(Figure~\ref{fig:lmc}) shows two bright sources near the Large
Magellanic Cloud, in the highly observed portion of the sky near the
south ecliptic pole. These sources are as bright as some of the
sources found by the \emph{WMAP} source-detection algorithm, and
indeed the brighter one is included in the subsequent catalogs of
\citet{Lopez-Caniego2007} and \citet{NieZhang2007}, based on
\emph{WMAP} data.  The presence of the nearby LMC may have interfered
with the source finding procedure.

\begin{figure}
\begin{center}
\includegraphics[width=\figwidth]{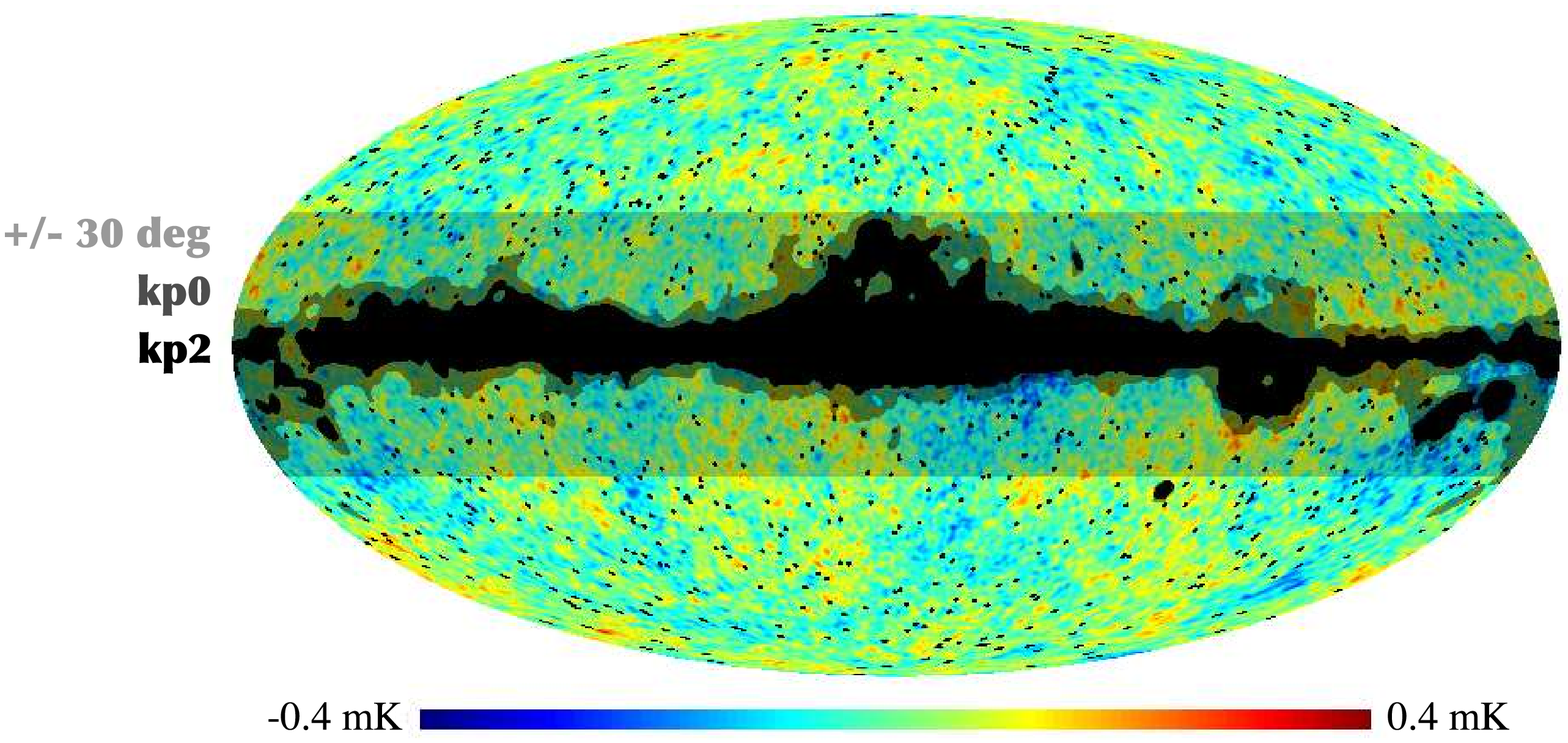}
\includegraphics[width=\figwidth]{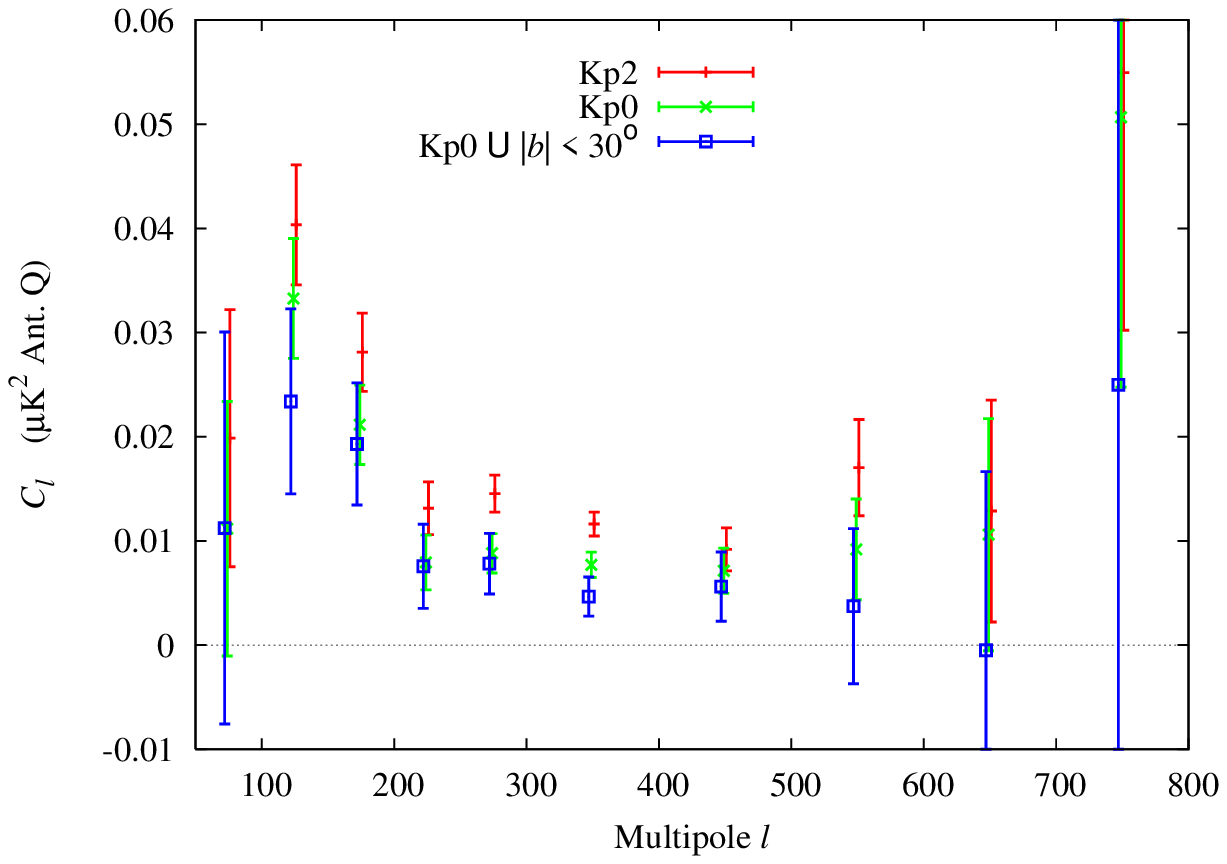}
\end{center}
\caption{(Top) Three masks used in this analysis.  The smallest, and
  most aggressive mask is the \emph{WMAP} Kp2 mask, intermediate is
  Kp0, and most conservative is the union of Kp0 and a $|b| <
  30^\circ$ cut.  (Bottom) The source power spectrum, computed with
  three different masks.  As the mask becomes larger, the power goes
  down, suggesting a concentration of unmasked source power near the
  galactic plane.}
\label{fig:maskdep}
\end{figure}

Changing the sky cut gives further indications of an anisotropic
source population.  In addition to the Kp2 cut used in the
cosmological analysis, we recomputed the point source fit using
cross-spectra generated with two additional masks and uniform
weighting: the Kp0 mask and a conservative mask consisting of the
union of Kp0 with a $|b| < 30^\circ$ cut galactic cut (see
Figure~\ref{fig:maskdep}).
As seen in Table \ref{tab:fits}, the source power drops significantly
as we expand the masks, indicating that \textit{unmasked} sources are
brighter or more common near the plane.  This raises two
possibilities. Either some of the sources are galactic in origin, or
sources are less efficiently found and masked near the plane.

\begin{figure}
\begin{center}
\includegraphics[width=\figwidth]{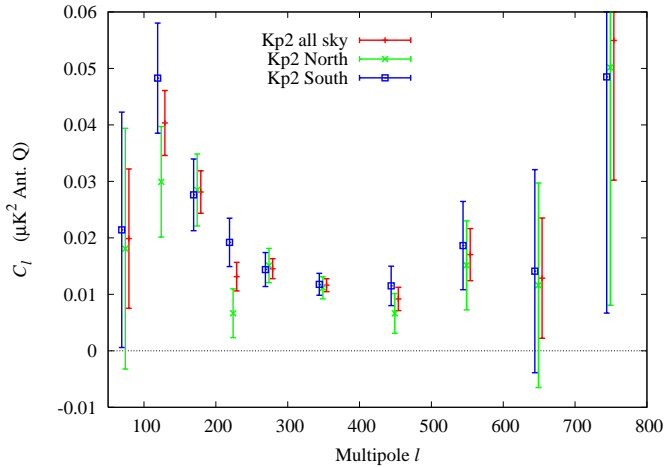}
\end{center}
\caption{The source power spectrum, computed in hemispheres with the Kp2 mask.}
\label{fig:hemi}
\end{figure}

From a visual inspection of the Q$-$ILC difference map, we find $\sim
6$ more bright point-like objects, many near the galactic plane, and
all in the southern hemisphere (see Table \ref{tab:new-sources}).
Many of these objects are also already noted in the
the \citet{Lopez-Caniego2007} and \citet{NieZhang2007} catalogs.
Computing the power spectrum in hemispheres (with the Kp2 mask), the source amplitudes straddle the value for the whole sky.  At $l<200$, the Northern hemisphere has slightly less source power and the the Southern slightly more  (See Figure~\ref{fig:hemi}).

The \emph{WMAP} source mask, built from a variety of catalogs
\citep{Hinshaw2003,Hinshaw2007}, may not cover the sky evenly, because
of differences in sensitivity in different regions. It may be more
difficult to identify sources amid large galactic foregrounds.  This
would be particularly true for flat spectrum sources in lower
frequency radio surveys.  The \emph{WMAP} source mask has noticeable
gaps near the galactic plane, but outside the Kp2 and Kp0 masks.

\begin{deluxetable}{ccccc}
\tablewidth{\figwidth} 
\tabletypesize{\small} 
\tablecaption{Sources found in  Q$-$ILC\label{tab:new-sources}}
\tablecolumns{5}
\tablehead{RA&Dec.&Gal. lon.&Gal. lat.&ID}
\tablecomments{Point-like objects, found visually in a Q$-$ILC difference map.  Marked sources (*,$^\dag$) were found respectively in the catalog of \citet{Lopez-Caniego2007} and \citet{NieZhang2007}.}
\label{tab:new-sources}

\startdata
00 43 14& -73 16 30 &
 303.7	&-44.0	 & PMNJ0047-7308*,$^\dag$\\ 
01 04 11& -72 08 02 &
 301.5	&-45.0 & PMNJ0059-7210*,$^\dag$ \\ 
04 51 56& -69 31 10 &
 276.2	&-33.7 & --- \\ 
05 20 56& -66 10 24 &
 281.0	&-35.6& PMNJ0506-6109*,$^\dag$ \\ 
06 49 57& -16 56 12 &
 227.9       &-8.0   & PMNJ0650-1637* \\ 
20 50 59& +28 51 35 &
 72.5	&-9.7 & ---$^\dag$ \\ 
20 52 02& +31 55 18 &
 75.1	&-8.0 & ---\\ 
20 57 54& +31 29 08 &
 75.6	&-9.2 & ---$^\dag$ \\ 
\enddata

\end{deluxetable}

\begin{figure}
\begin{center}
\includegraphics[width=\figwidth]{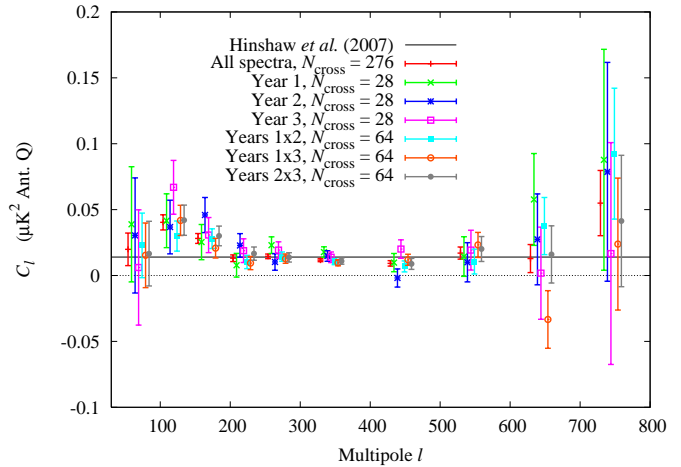}
\end{center}
\caption{Year by year comparisons of the point source power spectrum.  }
\label{fig:years}
\end{figure}

Two jackknife tests, breaking the data into subgroups, also show some
peculiarities.  First, we divided up the cross spectra based on the years of observation (Figure~\ref{fig:years}).
Each individual year accounts for 28 cross-spectra, while a pair of
years accounts for 64.  At $l > 200$, the fits for individual years
are always as greater than for pairs of different years.  The largest
estimate is the (year~3~$\times$~year~3) fit, at $A = 0.017 \pm 0.003$
$\mu$K$^2$, while the smallest is (year~1~$\times$~year~2), at $0.011 \pm
0.002$ $\mu$K$^2$.  This seems unlikely to be due to chance, and could
have a number of causes.  For example, a slight cross-correlation
between Q1 and Q2 or between Q and V, introducing a small noise bias
in the cross spectra, could have this effect.  Removing either Q1 or
Q2 from the source estimate, the amplitude drops to $A = 0.010 \pm
0.001$ $\mu$K$^2$.

\begin{figure}
\begin{center}
\includegraphics[width=\figwidth]{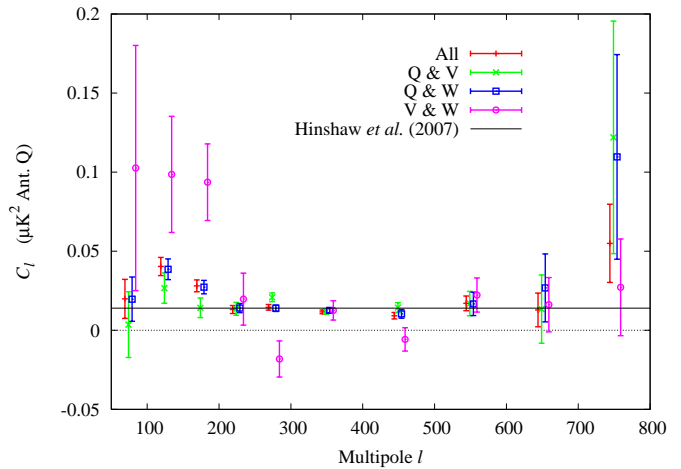}
\end{center}
\caption{The source power spectrum estimate for combinations of bands.}
\label{fig:bands}
\end{figure}

However, this dependence on the Q band could also be due to the shape
of the source spectrum.  For the second jackknife test, we broke the
cross spectra up by band.  We considered combinations with Q and V
bands only, Q and W only, and V and W only.  These spectra are shown
in Figure~\ref{fig:bands}.
At $l>200$, the V+W combinations are very noisy, and have the lowest
source fit, $A = 0.006 \pm 0.004$ $\mu$K$^2$, though the $\chi^2$ is
poor.  The Q+V combination has the largest source amplitude, $A =
0.016 \pm 0.001$ $\mu$K$^2$.  This discrepancy could mean that the
source spectrum is incorrect: If the true spectrum is steeper than the
$\beta = -2$ we have used, or the spectrum steepens between V and W,
one could observe this effect.  In this case, the Q+V combination
gives the most accurate measurement of the Q band source
contamination, but this is not used in the cosmological analysis.  To
get the correction in V and W bands, we are better off to scale from
the lower amplitude, although even this is not completely
satisfactory.  Using the wrong spectrum for sources means that
the shape of the source correction will be somewhat incorrect.
We gradually steepened the source spectrum, and re-fit.  The Q+V, Q+W,
and V+W fits can be brought within $1 \sigma$ by setting $\beta =
-2.5$.  These can be made equal (at $A=0.012$ $\mu$K$^2$) by setting
$\beta \sim -2.9$, which compared to $\beta = -2.0$ would represent a 50
percent smaller amplitude in V and an 80 percent smaller amplitude in
W.  If true, the source correction should be made correspondingly smaller.
(We continue to quote values of $A$ in Q-band assuming $\beta = -2.0$ scaling to
V and W.)

In the appendix, we discuss the impact of beam errors.  These tend to
have little impact on the spectrum at $l>200$, since the resulting CMB
leakage is large where the CMB is large, at lower $l$.

Finally, we must decide by how much to correct the combined
\emph{WMAP} spectrum.  There is a strong detection of unresolved power
in the \emph{WMAP} spectra, but as we have seen through the above tests, our
knowledge of the character of these sources is poor.  Our best fit
amplitudes including all the data are $A = 0.012$ $\mu$K$^2$ for flat
weighting and $A = 0.015$ $\mu$K$^2$ for $N_{\rm obs}$ weighting.  But
the spectral effects (from considering V and W without Q) indicate it
might be much lower, while individual year fits are much higher.  We
then are forced to the unsatisfying course of artificially inflating
the error bars beyond their nominal statistical values, in order to
account for these possibilities.  Therefore we set the error on the
source estimates at $\sigma_A = 0.005$ $\mu$K$^2$.


To obtain our final estimate for cosmological parameters, we correct
the spectrum for point sources using the two amplitudes quoted above
(uniform weights for $l<500$ and $N_{\rm obs}$ weights for $l>500$),
and our modified likelihood code.  The marginalized parameters from
the resulting Markov chains are given in Table \ref{tab:parameters},
both considering \emph{WMAP} data alone and including data from ACBAR \citep{kuo2004} 
and BOOMERanG \citep{jones2006,montroy2006,piacentini2006}. Because of
the new likelihood and the broadened error bars, this new value of
$n_s = 0.970\pm0.017$ is only $1.8\sigma$ away from 1.  Another consequence
is that $\sigma_8$ increases to $0.778 \pm 0.045$.

\begin{deluxetable}{lrr}
\tabletypesize{\small} 
\tablecaption{Cosmological parameters\label{tab:parameters}}
\tablecolumns{3}
\tablehead{Parameter & \emph{WMAP} &  Source/likelihood corrected }
\startdata
\cutinhead{\emph{WMAP} data only}
 $\Omega_{\textrm{b}}\,h^2$       & $0.0223\pm0.0007$ & $0.0220 \pm 0.0008$ \\ 
 $\Omega_{\textrm{m}}$            & $0.237\pm0.034$   & $0.242 \pm 0.037$    \\ 
 $h$                              & $0.735\pm0.032$   & $0.731 \pm 0.034$    \\ 
 $\tau$                           & $0.088\pm0.030$     & $0.090 \pm 0.030$    \\
 $n_{\textrm{s}}$                 & $0.951\pm0.016$   & $0.968 \pm 0.017$    \\ 
 $\sigma_8$                       & $0.742\pm0.051$   & $0.780 \pm 0.052$    \\
\cutinhead{\emph{WMAP} + ACBAR + BOOMERANG}
 $\Omega_{\textrm{b}}\,h^2$       & $0.0232\pm0.0007$ & $0.0224 \pm 0.0007$ \\ 
 $\Omega_{\textrm{m}}$            & $0.233\pm0.034$   & $0.234 \pm 0.032$    \\ 
 $h$                              & $0.739\pm0.033$   & $0.742 \pm 0.032$    \\ 
 $\tau$                           & $0.088\pm0.032$   & $0.092 \pm 0.030$    \\
 $n_{\textrm{s}}$                 & $0.951\pm0.016$   & $0.970 \pm 0.017$    \\ 
 $\sigma_8$                       & $0.739\pm0.051$   & $0.778 \pm 0.045$    \\
\enddata
\tablecomments{Comparison of marginalized parameter results obtained
  from Table 5 of \citet{Spergel2007} (second column) and 
after our modifications to the source correction and likelihood (third column).}
\end{deluxetable}

\subsection{Excess at $l < 200$}
\label{sec:lowl_excess}

We now turn to the $l < 200$ feature, which is inconsistent with a
white noise spectrum.  This is present in both the uniform and $N_{\rm
  obs}$ weighted spectra, and in each year and pair of years.  Because
of its shape and prominence at low $l$, we initially considered two
specific explanations. The first possibility concerned mis-estimation
of the overall multiplicative map calibration of each DA map: If two
maps are calibrated differently, the weights $W_L^{ij}$ in Equation
\ref{eq:amplitude} would not cancel the CMB component precisely. Thus,
one would observe a leakage from the CMB signal into the point source
spectrum, with a signature resembling the CMB power spectrum. A
similar effect would be caused by beam uncertainties. In Appendix A,
we present the formalism to take these uncertainties into account in
the method described in Section \ref{sec:method}, and the results from
the corresponding analysis of the \emph{WMAP} data are presented in the
bottom Section of Table \ref{tab:fits}. The conclusion from these
computations is that neither calibration nor beam uncertainties can
explain this effect.

A third hypothesis is residual galactic foregrounds, which should show
through the mask and frequency dependency.  As the galactic cut widens
from Kp2 to the wide cut (Figure~\ref{fig:maskdep}), the bin from
$l=100$--$150$ drops about 0.017 $\mu$K$^2$.  At the same time the fit
for the white noise level drops by 0.006 $\mu$K$^2$.  Subtracting off
the white noise levels for each, the component in the excess has
dropped by about 40 percent.  The power in the excess is still
significant, even for this broad cut.

Next, we turn to the estimates using only two bands.  In this case, we
see that the feature is strongly enhanced in the V+W combinations,
about the same in the Q+W combination, but clearly diminished in the
Q+V combination.  The latter appears consistent with the flat source
spectrum.  This may indicate that the excess is associated with the W
band. On the other hand, because the W band spectra tend to carry
negative weight in these estimates (Figure \ref{fig:weights}), this
excess would represent a deficit of power in W, which is peculiar.  It
may indicate an over-subtraction of the galactic foreground template
in W band, which could have consequences for the cosmological
analysis.  At this point, it is difficult to be definitive.

\section{Conclusions}
\label{sec:conclusions}

We have reanalyzed the source correction procedure for the three-year
data release of \emph{WMAP}.  First, we considered the impact of this
procedure in the \emph{WMAP} likelihood code. Surprisingly, we found that the
\emph{WMAP} likelihood does not react to changes in the point source
correction error.  We have devised a modified likelihood which does
respond as expected, although we note that more work is needed to
completely validate this approach, as it couples to the important
problem of how to approximate a non-Gaussian likelihood with fitting
formulas over a wide multipole range.  To conclude $n_s < 1$, a
precision measurement of the source contamination is required.  We
note that the modes not contaminated favor $n_s$ consistent with
unity.

Second, we found several indications that the unmasked source
population in \emph{WMAP} data is anisotropic.  This implies that the
combined spectrum should be corrected differently in two multipole
regions, based on the weighting of the map.  Anisotropy in the
unmasked sources is unexpected, but can be turned to an advantage: By
very carefully masking near the ecliptic poles and galaxy, or
employing a wide galactic cut, the point source contamination can be
cut substantially.  This gain must be weighed against the reduction of
the sky area.

We note irregularities in jackknife tests of the source fit, grouped
by time of observation or frequency band.  This prompted us to adopt
large errors on the source fit, to account for systematics beyond the
estimate of statistical error which accompanies our measurement.  This
step is necessary to treat the source correction conservatively.  With
the modified likelihood and reduced source amplitude from ignoring the
excess at $l<200$, these enlarged error bars are responsible for
raising our values of $n_s$ and $\sigma_8$.

Finally, the previously noted anomalous $l < 200$ feature is still
present, shows signs of being spatially associated with the galaxy,
and is most strongly associated with the W band.  On the other hand,
it does not appear to be associated with calibration or beam errors.
It may represent an over-subtraction of the foreground template in W,
although further investigation is warranted. However, the immediate
conclusion is that this part of the spectrum should not be used to
infer the point source amplitude at higher $l$'s.

\begin{acknowledgements}
  We thank the \emph{WMAP} team for useful discussions and for
  providing additional data.  In particular we thank Gary Hinshaw,
  Michael Nolta, and Lyman Page, who suggested examining the effect of
  beam uncertainties.  HEALPix software \citep{healpix} was used to
  deriving some results in this paper. We also acknowledge use of the
  Legacy Archive for Microwave Background Data Analysis (LAMBDA).  HKE
  acknowledges financial support from the Research Council of
  Norway. 
  This work was partially performed at the Jet Propulsion Laboratory,
  California Institute of Technology, under a contract with NASA.
\end{acknowledgements}

\bibliographystyle{hapj}  
\bibliography{wmap_ns_certainty}

\appendix
\section{Leakage from Calibration and Beam Errors}

In the appendix, we estimate the leakage from CMB into the point
source estimate.  We consider two sources of leakage, the overall
calibration of the map and the errors in the beam deconvolution.  Each
of these can cause the CMB spectrum to have a slightly different
amplitude or shape in each of the cross spectra.  This means that the
CMB will not cancel itself in the weighted linear combination to give
the source estimate.  This leakage can be accounted for in the cross
spectrum covariance.  The calibration is a constant function of $l$, and the beam error is nearly so over a wide range of scales, so the shape of the leakage term strongly follows the CMB power spectrum, which drops rapidly with $l$.  

The calibration uncertainty is simpler, so we begin there.  In the signal (s) dominated regime, without including point sources, we can model the calibration errors with
\begin{equation}
  d_i = (1+g_i) s.
\end{equation}
where the calibration in map $d_i$ is quantified with dimensionless factors $g_i$ with $\langle g_i \rangle = 0$ and $\langle g_i g_j \rangle = \sigma_g^2 \delta_{ij}$.  Then the estimated cross-spectra are
\begin{equation}
  D_l^{ij} = (1 + g_i + g_j + g_i g_j) \hat C_l.
\end{equation}
Where $\hat C_l$ is the power spectrum of the particular sky realization (although in the actual calculation, we substitute the \emph{WMAP} best-fit spectrum).
In the cross spectrum  $i \neq j $, so this is an unbiased estimator, $\langle  D_l^{ij} \rangle = \hat C_l$.  
The covariance is
\begin{equation}
{\rm Cov}(D^{ij}_l, D^{pq}_{l'}) = \hat C_l \hat C_{l'} \left[ \sigma^2_g (\delta_{ip} + \delta_{iq} + \delta_{jp} + \delta_{jq}) 
 + \sigma_g^4 (\delta_{ip}\delta_{jq} + \delta_{iq}\delta_{jp}) \right].
\end{equation}
Essentially, the variance only gets a contribution in terms where the two cross-spectra share a map.

The point source estimate is
\begin{equation}
A_l = \sum_{l'} \sum_{(ij)} W_{ll'}^{ij}  D_{l'}^{ij}  = \sum_{l'} \hat C_{l'} \sum_{(ij)} W_{ll'}^{ij} (g_i + g_j + g_i g_j)
\end{equation}
where one term has dropped out because the weights sum to zero.  Because we have not included any source contribution, we have $\langle A_l \rangle = 0$, where the ensemble average is over the calibration factors.
For plotting, its useful to compute the variance due to the calibration factors
\begin{eqnarray}
\langle A^2_l \rangle &=& \sum_{l'l''} \hat C_{l'} \hat C_{l''}  \sum_{(ij)\,(pq)}  W_{ll'}^{ij}  W_{ll''}^{pq} \left[ \sigma^2_g (\delta_{ip} + \delta_{iq} + \delta_{jp} + \delta_{jq})  + \sigma_g^4 (\delta_{ip}\delta_{jq} + \delta_{iq}\delta_{jp}) \right]
\label{eqn:calvar}
\end{eqnarray}

A word about binning the spectrum is appropriate here.  In practice we bin the spectrum because the large number of cross-spectra (276) and multipoles ($\sim 800$) slows the computation, and the signal-to-noise per multipole in low.  We may
define a binning matrix $G_{Ll}$ which averages quantities in non-overlapping bands indexed by $L$, and compute our estimate from the binned cross spectra $\sum_l G_{Ll}   D_l^{ij}$.
The quantity we are now computing is the variance in a bin.  Since the weights are also computed in bands, this reduces to replacing the $l$s with $L$s and replacing each power spectrum term $C_l$ with $\sum_l G_{Ll} \hat C_l $.

In \citet{Jarosik2007}, the calibration uncertainty is quoted as 0.5 percent, so $\sigma_g = 0.005$.  As written, we are assuming the all maps are uncorrelated, the case if the calibration fluctuations are dominated by noise as expected \citep{Page2007Personal}. In this case, fluctuations in the calibration are fairly small compared to the source estimate.  If the calibration is dominated by something else, say foregrounds, then whole bands could be correlated, and the calibration leakage can be substantial compared to the source fit.

In a similar way, we can estimate the beam uncertainties.  We begin with
\begin{equation}
d^i_{lm} = \hat B^i_l ( 1 + E^i_l ) s_{lm},
\end{equation}
where $\hat B^i_l$ is an unbiased measurement of the beam, and $E^i_l$ is the (small) fractional difference between the true beam and the measured beam.  We assume all beams have independent errors, and define a $\delta$-function-like object $\delta^B_{ij}$, which is unity when maps $i$ and $j$ share a beam, and zero otherwise.  With this definition, we may use a describe the beam error with a Gaussian distribution with $\langle E^i_l \rangle = 0$ and 
\begin{equation}
\langle E^i_l E^j_l \rangle = \Sigma^{B,i}_{ll'} \delta^B_{ij}.
\end{equation}
That is, the beam errors are correlated in $l$ but not between beams.  The beam-deconvolved cross spectra are 
\begin{equation}
D^{ij}_l = (1 +  E^i_l)(1 + E^j_l) \hat C_l.
\end{equation}
Following the same procedure as with gain calibrations, we find a similar expression for the covariance:
\begin{equation}
 {\rm Cov}(D^{ij}_l, D^{pq}_{l'}) =  \hat C_l \left[   \Sigma^{B,i}_{ll'} (\delta^B_{ip} + \delta^B_{iq})
+  \Sigma^{B,j}_{ll'} (\delta^B_{jp} + \delta^B_{jq})  
+  \Sigma^{B,i}_{ll'} \Sigma^{B,j}_{ll'} (\delta^B_{ip}\delta^B_{jq} + \delta^B_{iq}\delta^B_{jp}) \right] \hat C_{l'}. 
\end{equation}


\emph{WMAP} found that the beam uncertainty can be well represented as a small number of orthogonal modes  ($\sim 10$) \cite{Hinshaw2003,Hinshaw2007}:
\begin{equation}
\Sigma^{B,i}_{ll'} = \sum_r U^i_{rl} U^{i}_{rl'}.
\end{equation}
where $r$ denotes the mode.
Here the binning is a little more complicated than for the calibration uncertainty.  Instead of simply $\hat C_l$, the quantities which must be binned are $\hat C_l U^i_l$ (for the first-order terms) and $\hat C_l U^i_{rl} U^j_{tl}$ (for the second order term).

\begin{figure}
\begin{center}
\includegraphics[width=0.98\figwidth]{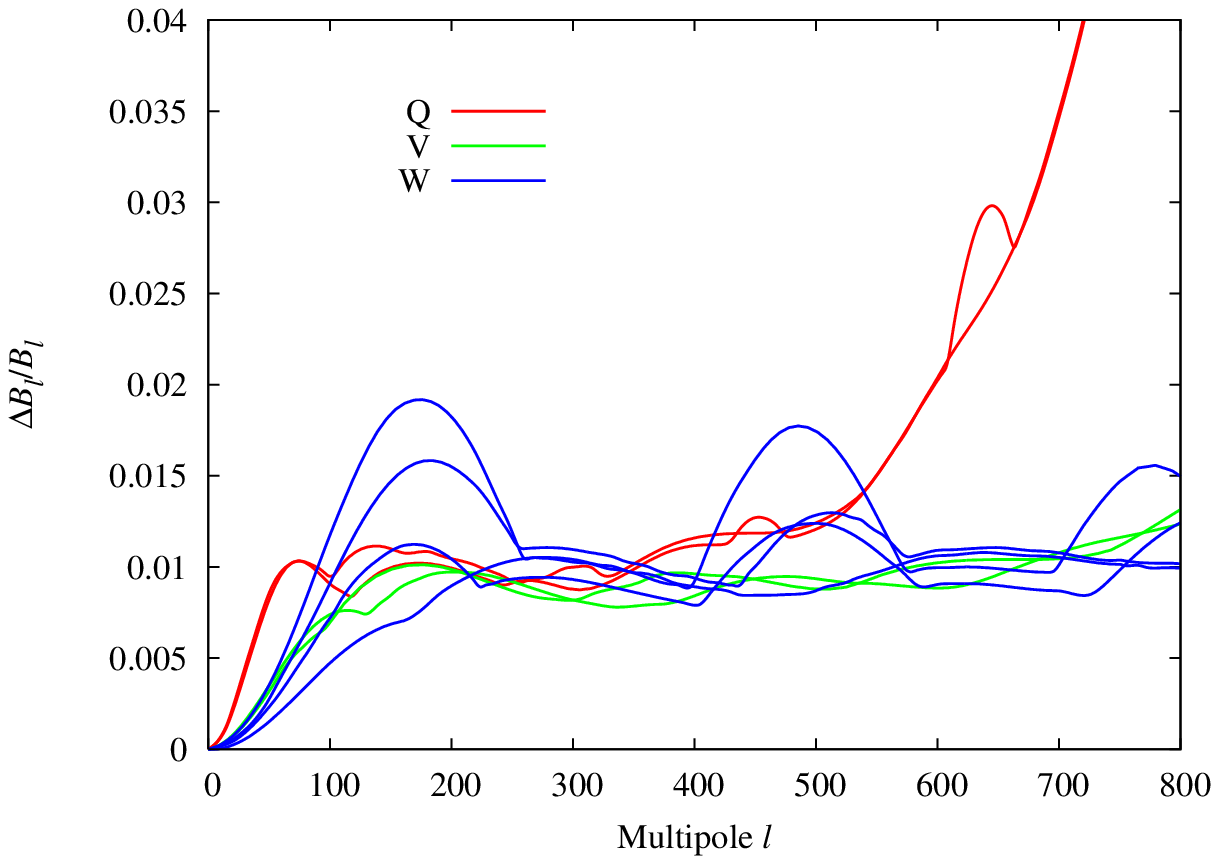}
\includegraphics[width=0.98\figwidth]{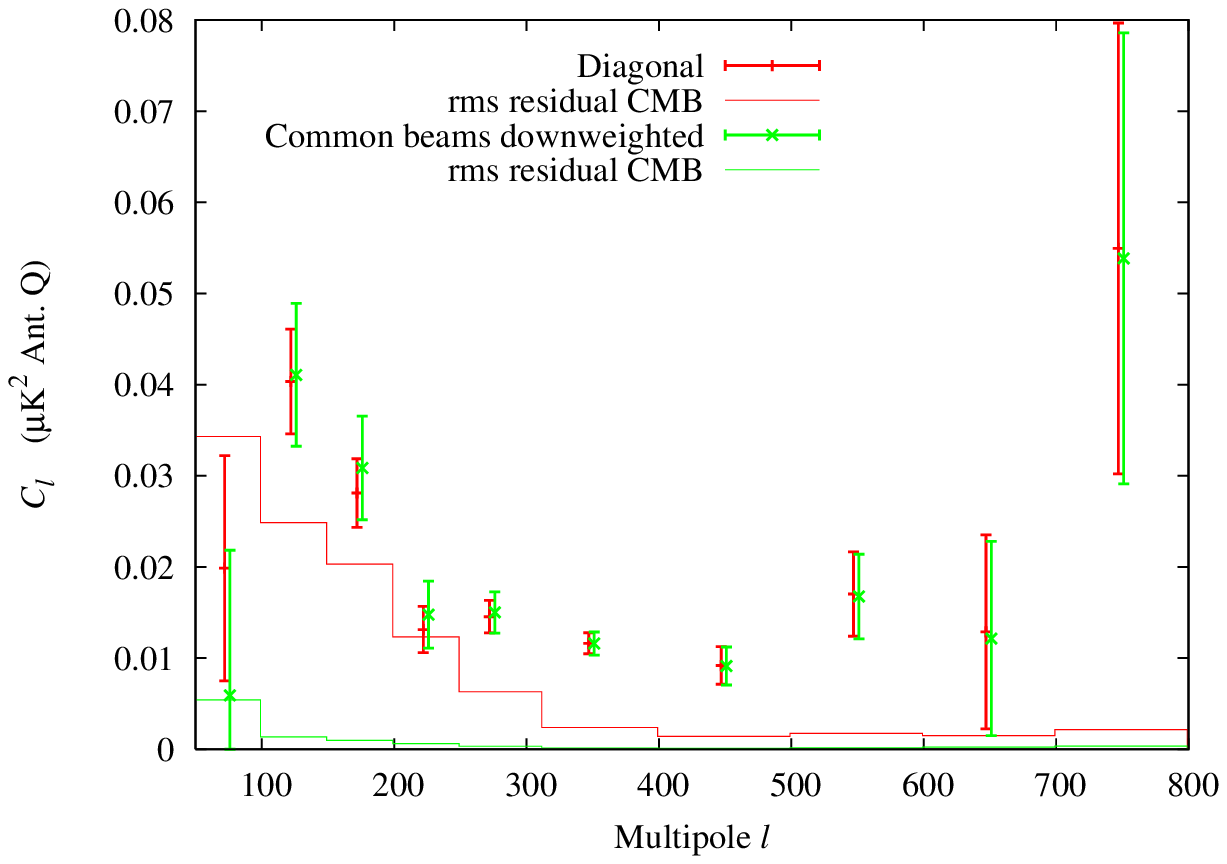}
\end{center}
\caption{(Left) Beam errors taken from \citet{Jarosik2007}.  The Q-band errors rise smoothly to $\sim 0.08$ at $l=800$. (Right) rms leakage of the CMB into the point source measurement, with and without taking the contribution to the beam covariance into account.  }
\label{fig:beamerr_fit}
\end{figure}

At this time, the beam modes for each \emph{WMAP} differencing assembly are not public.  (Only the modes for the combined spectrum are included in the likelihood code.)  To approximate the beam modes, we take the beam errors from \citet{Jarosik2007} (shown in Figure~\ref{fig:beamerr_fit}).  We treat each of these as a single mode for the associated beam, correlating all the multipoles in that beam.  This should provide a conservative estimate of the beam covariance, though it does not capture all of its properties.
With these approximate modes, we can compute the beam variance and the rms CMB leakage for any set of weights.  This is plotted in the right panel of Figure \ref{fig:beamerr_fit}.  For our standard diagonal covariance, the rms beam leakage at first looks like a promising explanation for the $l < 200$ excess.  It has a similar shape, and strong correlations bin-to-bin.  However, when we recompute the weights taking the beam covariance into account, we produce a similar estimate for point sources in a combination which allows very little CMB leakage.  Given the estimated size of the rms residual CMB, it is surprising that the source estimates are so similar.  Perhaps this is an indication that the beam covariance is overestimated.  In any case, the $l < 200$ is either not due to beams, or the approximation for the beam modes is very poor.

\end{document}